\documentclass[aps,prl,reprint,superscriptaddress,10pt,a4paper,preprintnumbers,nofootinbib]{revtex4-1}
\usepackage{amsmath,amssymb,latexsym}
\usepackage{bm}
\usepackage{graphicx}
\usepackage{hepunits}
\usepackage[svgnames]{xcolor}

\allowdisplaybreaks

\begin{document}

\title{Two-loop triangle integrals with 4 scales for the $HZV$ vertex}

\author{Yuxuan Wang}
\email{wangyuxuan119@pku.edu.cn}
\author{Xiaofeng Xu}
\email{xuxiaofeng@pku.edu.cn}
\affiliation{School of Physics and State Key Laboratory of Nuclear Physics and Technology, Peking University, Beijing 100871, China}
\author{Li Lin Yang}
\email{yanglilin@pku.edu.cn}
\affiliation{School of Physics and State Key Laboratory of Nuclear Physics and Technology, Peking University, Beijing 100871, China}
\affiliation{Collaborative Innovation Center of Quantum Matter, Beijing, China}
\affiliation{Center for High Energy Physics, Peking University, Beijing 100871, China}

\begin{abstract}

We calculate analytically the two-loop triangle integrals entering the $\mathcal{O}(\alpha\alpha_s)$ corrections to the $HZV$ vertex with $V=Z^*,\gamma^*$ using the method of differential equations. Our result provides a prototype to study the analytic properties of multi-loop multi-scale Feynman integrals, and also allows fast numeric evaluation for phenomenological studies. We apply our results to the leptonic decay of the Higgs boson and to $ZH$ production at electron-positron colliders. Besides the top quark loop, we include also the bottom quark loop contributions, whose evaluation takes a lot of time using purely numeric methods, but is very efficient with our analytic results.

\end{abstract}


\maketitle

\section{Introduction}

The $HZV$ vertex, where $V=Z^*,\gamma^*$, is relevant for several important processes in Higgs physics, such as the leptonic decay $H \to Z^*Z^* \to 4l$ which is crucial in the discovery of the Higgs boson and is also important for measuring its mass and width. The vertex also enters the production of the Higgs boson through vector boson fusion ($ZZ \to H$) at the Large Hadron Collider (LHC), as well as the associate production of the Higgs boson with a $Z$ boson ($pp \to ZH$) at the LHC and at future Higgs factories ($e^+e^- \to ZH$). Theoretically, the $HZV$ vertex probes the electroweak symmetry breaking mechanism of the standard model, and is sensitive to alternative models of the Higgs sector. For example, if the Higgs is a composite Goldstone boson, the tree-level $HZZ$ coupling is modified by $\mathcal{O}(v^2/f^2)$ which can be as large as a few percent if the symmetry breaking scale $f$ is in the TeV range. Therefore, precise knowledge of the $HZV$ vertex is essential for the quest to better understand the properties of the Higgs boson and the mechanism of electroweak phase transition.

Experimentally, the $HZZ$ vertex can be measured with an uncertainty of 1.5\% at the High Luminosity LHC (HL-LHC) with an integrated luminosity of $\unit{3000}{\invfb}$~\cite{Cepeda:2019klc} (in the $\kappa$-interpretation). At future Higgs factories such as the Circular Electron Positron Collider (CEPC), the Future Circular Collider (FCC-ee) and the International Linear Collider (ILC), the experimental precision for the $HZZ$ vertex can reach the sub-percent level to about 0.2\%~\cite{Ruan:2014xxa, dEnterria:2016fpc, CEPCStudyGroup:2018ghi}.

\begin{figure}[h!]
\includegraphics[width=0.3\textwidth]{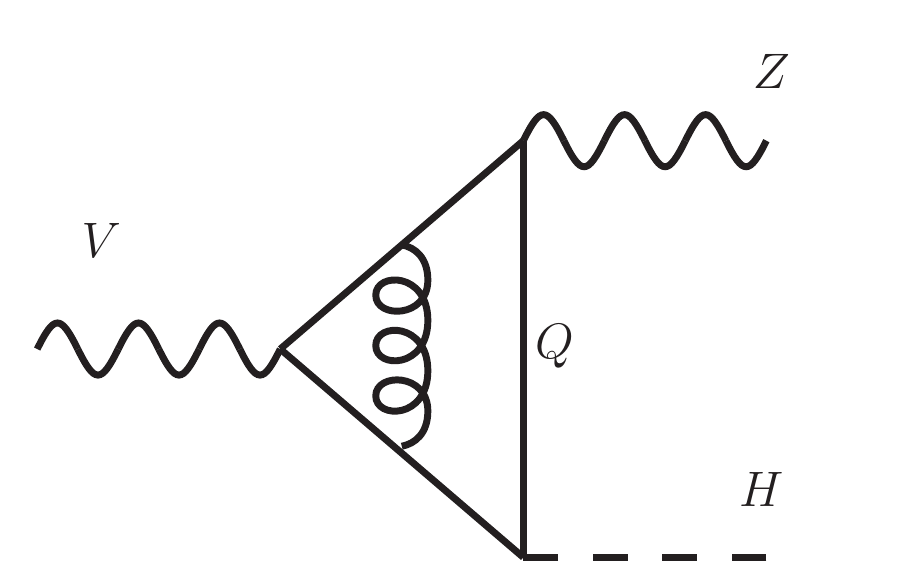}
\caption{\label{fig:dia}A typical diagram contributing to the $HZV$ vertex at $\mathcal{O}(\alpha\alpha_s)$. The heavy quark $Q$ can be the top quark $t$ or the bottom quark $b$.}
\end{figure}

In order to match the accuracy of future experimental measurements, it is necessary to provide high precision theoretical calculations for the $HZV$ vertex within the standard model perturbation theory. The next-to-leading order $\mathcal{O}(\alpha)$ corrections were given in \cite{Kniehl:1991hk, Denner:1992bc}. At the two-loop level, we are concerned with  the $\mathcal{O}(\alpha\alpha_s)$ mixed QCD-electroweak corrections. A typical diagram is depicted in Fig.~\ref{fig:dia}, where the heavy quark $Q$ running in the loop can be the top quark $t$ or the bottom quark $b$. To fix the notation, we consider the momenta assignments as $V(q) \to Z(p_Z) + H(p_H)$. The integrals we need to calculate are of the form
\begin{equation}
G_{\{a_i\}} \equiv m_Q^{4\epsilon} \int \frac{d^dk_1}{i\pi^{d/2}\Gamma(\epsilon)} \frac{d^dk_2}{i\pi^{d/2}\Gamma(\epsilon)} \prod_{i=1}^7 \frac{1}{D_i^{a_i}} \, ,
\end{equation}
where $d=4-2\epsilon$ in dimensional regularization, $k_1$ and $k_2$ are loop momenta, $\{a_i\} \equiv \{a_1,a_2,a_3,a_4,a_5,a_6,a_7\}$ are the powers of the propagators $D_i$ in the integrand, and the propagators are given by
\begin{align*}
&\{ k_1 - m_Q^2 \, , k_2^2 - m_Q^2 \, , (k_1-k_2)^2 \, , (k_1-p_Z)^2 - m_Q^2 \, ,
\\
&(k_1+p_H)^2 - m_Q^2 \, , (k_2-p_Z)^2 - m_Q^2 \, , (k_2+p_H)^2 - m_Q^2 \} \, .
\end{align*}
The results of the integrals are functions of the 4 Lorentz invariants $m_Q^2$, $q^2$, $p_Z^2$ and $p_H^2$.
We employ the method of integration-by-parts to reduce all these integrals into 41 master integrals, using the program packages \texttt{FIRE5}~\cite{Smirnov:2014hma}, \texttt{LiteRed}~\cite{Lee:2013mka} and \texttt{Reduze2}~\cite{vonManteuffel:2012np}. These master integrals can be computed numerically using the method of sector decomposition~\cite{Binoth:2000ps}, as performed in \cite{Gong:2016jys, Sun:2016bel}. Note that the numeric integration is highly resource-demanding if any of $q^2$, $p_Z^2$ and $p_H^2$ is larger than $4m_Q^2$, which is the case if one considers collider energies above the $t\bar{t}$ threshold, or if one wants to take into account the contribution from bottom quark loops.

In the case where $q^2, p_Z^2, p_H^2 < 4m_Q^2$, one may perform a series expansion of the integrals in $1/m_Q^2$. This approach has been taken in \cite{Gong:2016jys}. The benefit of this method is that after the expansion (which can be done at the amplitude level), the remaining integrals are single-scale ones which can be easily evaluated to analytic expressions. Therefore, one can implement the result straightforwardly into any event generators for phenomenological analyses.

If $q^2 > 4m_Q^2$ but $p_Z^2, p_H^2 < 4m_Q^2$, the $1/m_Q^2$ expansion fails but one may instead employ an expansion in powers of $p_Z^2$ and $p_H^2$. This is in spirit very similar to the method of Ref.~\cite{Xu:2018eos}, where a small $m_H$ expansion is performed for Higgs boson pair production at the LHC. This kind of expansion is generically valid for top quark loops, but will not work for bottom quark loops.

Given the disadvantage of the purely numeric method and the limited applicability of various approximations, the goal of this paper is to provide an exact analytic solution to the master integrals appearing in the $\mathcal{O}(\alpha\alpha_s)$ corrections to the $HZV$ vertex. The analytic expressions are valid for arbitrary values of the internal mass and the external momenta. This will allow fast numeric evaluations at any phase-space point, and will also serve as a prototype for analyzing the structure of loop integrals with many scales.

\section{Analytic solution}

To obtain the analytic solution for the master integrals, we employ the method of differential equations~\cite{Kotikov:1990kg, Kotikov:1991pm}. We define the dimensionless variables
\begin{equation}
x = -\frac{q^2}{4m_Q^2} \, , \quad y = -\frac{p_Z^2}{4m_Q^2} \, , \quad z = -\frac{p_H^2}{4m_Q^2} \, .
\end{equation}
We are able to find a basis $\vec{f}(x,y,z,\epsilon)$ of the master integrals which satisfies the canonical-form differential equation~\cite{Henn:2013pwa}
\begin{align}
d\vec{f}(x,y,z,\epsilon) &= \epsilon \, dA(x,y,z) \, \vec{f}(x,y,z;\epsilon) \nonumber
\\
&= \epsilon \sum_{i} A_i \, d\log(\alpha_i) \, \vec{f}(x,y,z,\epsilon) \, ,
\label{eq:deq1}
\end{align}
where $A_i$ are constant matrices independent of kinematic variables and the dimensional regulator. The ``letters'' $\alpha_i \equiv \alpha_i(x,y,z)$ are rational functions of $x$, $y$, $z$ and the following 4 kinds of square roots:
\begin{gather}
R_1(x) \equiv \sqrt{x(x+1)} \, , \quad R_1(y) \, , \quad R_1(z) \, , \nonumber
\\
R_2 \equiv R_2(x,y,z) \equiv \sqrt{\lambda(x,y,z)} \, .
\label{eq:sqrt}
\end{gather}
with the Kallen function given by
\begin{equation}
\lambda(x,y,z) \equiv x^2 + y^2 + z^2 - 2xy - 2yz - 2zx \, .
\end{equation}
The solution to the canonical-form differential equation \eqref{eq:deq1} can be formally written as a path-ordered integral from the boundary point $\vec{r}_0 = (x_0,y_0,z_0)$ to the point $\vec{r} = (x,y,z)$
\begin{equation}
\vec{f}(x,y,z;\epsilon) = \mathcal{P} \exp \bigg( \epsilon \int_{\vec{r}_0}^{\vec{r}} dA \bigg) \vec{f}(x_0,y_0,z_0;\epsilon) \, .
\label{eq:sol1}
\end{equation}
We choose the boundary point to be $\vec{r}_0 = (0,0,0)$, where the values of the master integrals are simply given by
\begin{equation}
f_i(0,0,0;\epsilon) = \delta_{i1} \, .
\label{eq:boundary}
\end{equation}
In practice, we are interested in the master integrals as power series in $\epsilon$
\begin{equation}
\vec{f}(x,y,z;\epsilon) = \sum_{n=0}^\infty \vec{f}^{(n)}(x,y,z) \epsilon^n \, ,
\end{equation}
where the $n$-th order coefficient function with $n > 0$ can be represented as a linear combination of Chen's iterated integrals~\cite{Chen:1977oja} with transcendental weight $n$ of the form
\begin{equation}
\int_{\vec{r}_0}^{\vec{r}} d\log(\alpha_{i_n}(\vec{r}_n)) \cdots \int_{\vec{r}_0}^{\vec{r}_3} d\log(\alpha_{i_2}(\vec{r}_2)) \int_{\vec{r}_0}^{\vec{r}_2} d\log(\alpha_{i_1}(\vec{r}_1)) \, .
\label{eq:iterated}
\end{equation}
These integrals can be mapped to the symbol representation
\begin{equation}
\alpha_{i_1}(\vec{r}) \otimes \alpha_{i_2}(\vec{r}) \otimes \cdots \otimes \alpha_{i_n}(\vec{r}) \, ,
\end{equation}
which can be manipulated using its algebraic properties.
In general, these iterated integrals can be converted to linear combinations of multiple polylogarithms (MPLs)~\cite{Goncharov:1998kja}. If all the letters are rational functions, this conversion can be handled straightforwardly, at least for low weights. However, in our case the presence of the 4 kinds of square roots in Eq.~\eqref{eq:sqrt} makes the conversion much more difficult. Therefore, we need to describe the procedure in more detail.

Only 3 letters appear at weight 1, which are $\beta_i \equiv \beta(r_i)$ with $r_1 = x$, $r_2 = y$, $r_3 = z$ and
\begin{equation}
\beta(t) \equiv \frac{R_1(t)-t}{R_1(t)+t} \, .
\label{eq:beta}
\end{equation}
The inversion of the above formula gives
\begin{equation}
t = \frac{\big( 1 - \beta(t) \big)^2}{4\beta(t)} \, , \quad R_1(t) = \frac{1-\beta^2(t)}{4\beta(t)} \, .
\label{eq:R1t}
\end{equation}
Noting that $\beta(t) = 1$ for $t = 0$, we conclude that the weight-1 part of the master integrals can be entirely written in terms of $\log(\beta(x))$, $\log(\beta(y))$ and $\log(\beta(z))$ which satisfy the right boundary condition \eqref{eq:boundary}. At weight 2, another 13 letters come into play, and the independent integrable symbol tensors can be chosen as
\begin{gather}
\beta_i \otimes \beta_i \, , \quad
\beta_i \otimes \beta_j + \beta_j \otimes \beta_i \, , \quad
\beta_i \otimes r_i \, , \quad
\beta_i \otimes (1+r_i) \, , \nonumber
\\
\frac{\beta_i \beta_j}{\beta_k} \otimes \bigg( 1 - \frac{\beta_i \beta_j}{\beta_k} \bigg) \, , \quad
(\beta_i\beta_j\beta_k) \otimes ( 1 - \beta_i\beta_j\beta_k ) \, , \nonumber
\\
\beta(x) \otimes \frac{x(x-y-z) - R_1(x) R_2}{x(x-y-z) + R_1(x) R_2} + (x \leftrightarrow y) + (x \leftrightarrow z) \, ,
\label{eq:weight2}
\end{gather}
where $i$, $j$ and $k$ take distinct values from 1, 2 and 3. Except for the last one, the above symbol tensors can be converted to MPLs with the help of the known symbol maps of logarithm and dilogarithm functions.

The last symbol tensor in Eq.~\eqref{eq:weight2} involves all 4 kinds of square roots, which cannot be rationalized simultaneously. To derive its functional form, we exploit its iterated integral representation. Being an integrable symbol tensor, the corresponding combination of integrals is path-independent. We can therefore choose a specific path to evaluate the 3 terms separately, while keeping in mind that only the sum of them is meaningful (i.e., we should choose the same path for all 3 terms). We parametrize the path as $t\vec{r}$, with $t$ going from 0 to 1. As an example, the first term can be written as
\begin{equation}
I_1 = \int_{t=0}^{t=1} \log(\beta(tx)) \, d\log \frac{tx(x-y-z) - R_1(tx)R_2}{tx(x-y-z) + R_1(tx)R_2}\, ,
\end{equation}
where we have exploited the fact that $\lambda(x,y,z)$ is a homogeneous quadratic polynomial of $x$, $y$ and $z$, so that
\begin{equation}
R_2(tx,ty,tz) = t R_2(x,y,z) = t R_2 \, .
\end{equation}
According to Eq.~\eqref{eq:R1t}, the square root $R_1(tx)$ can be rationalized via a change of variable to $u \equiv 1-\beta(tx)$, and the integral $I_1$ is then given by
\begin{equation}
I_1 = \int_{u=0}^{u=1-\beta(x)} G(1;u) \, d\log \frac{u(x-y-z+R_2) - 2R_2}{u(x-y-z-R_2) + 2R_2} \, .
\end{equation}
Note that in the above formula, $x$, $y$ and $z$ should be treated as constants when taking the differential (i.e., only $u$ is the active variable). The integration over $u$ can be performed using the definition of MPLs, and the first term then becomes
\begin{align}
I_1 &= G \bigg( \frac{2R_2}{R_2+x-y-z}, 1; 1 - \beta(x) \bigg) \nonumber
\\
&- G \bigg( \frac{2R_2}{R_2-x+y+z}, 1; 1 - \beta(x) \bigg) \, ,
\end{align}
and the functional form of the whole symbol tensor can be obtained via permutation.

We now turn to the weight-3 part of the solution. With the presence of non-rational functions, it is a highly complicated task to determine integrable symbol tensors at high weights. Nevertheless, since the master integrals as a whole correspond to a path-independent combinations of iterated integrals, we can still choose a specific path to evaluate them. At weight 3, there can be two $R_1$ functions appearing together with $R_2$ in one symbol. Without loss of generality, we take them to be $R_1(x)$ and $R_1(y)$. Similar to the weight-2 case, we choose the path parameterized by $t\vec{r}$. To obtain the solution in terms of MPLs, we now need to rationalize $R_1(tx)$ and $R_1(ty)$ simultaneously, which can be done by the change of variable
\begin{equation}
t = \frac{v^2(2+v)^2}{4(1+v) \left( (1+v)\sqrt{x}+\sqrt{y} \right) \left( (1+v)\sqrt{y} + \sqrt{x} \right)} \, .
\label{eq:tv}
\end{equation}
The integration over $v$ can then be handled using the definition of MPLs. Note that this variable change may lead to a proliferation of terms, and should only be used when necessary. 

Using the above method, we are able to express all the master integrals in analytic form in terms of MPLs up to weight-3. Note that it is very rare that such two-loop integrals with many scales can be solved in closed form, and our result serves as a prototype to study the analytic structure of multi-loop multi-scale Feynman integrals. For example, it is interesting to investigate the behaviors of the scattering amplitude in various asymptotic limits, such as the massless limit and the threshold limit. These will be presented in a forthcoming article~\cite{future}.

Finally, in the weight-4 part of the solution, all the 4 square roots in Eq.~\eqref{eq:sqrt} can appear in a single symbol. We are not able to simultaneously rationalize $R_1(tx)$, $R_1(ty)$ and $R_1(tz)$ via a variable change (with respect to $t$). It is therefore not possible to convert the symbols to MPLs using the above method. It remains possible that one can construct the solution in terms of polylogarithms, using the function arguments at weight-2 and weight-3 as hints. We leave this for future investigation. In the current work, we evaluate the weight-4 part as a one-fold numeric integral. This is sufficiently fast for practical purposes. 

\section{Numeric results}

We now use our analytic result to calculate the $\mathcal{O}(\alpha\alpha_s)$ corrections to the $e^+e^- \to ZH$ production cross section and the $H \to ZZ^* \to 4l$ partial width. We consider both the top quark and the bottom quark in the loop.
In our numeric calculations, we take $m_t = \unit{173.3}{\GeV}$, $\Gamma_t = \unit{1.35}{\GeV}$,  $m_b = \unit{4.78}{\GeV}$, $m_Z = \unit{91.1876}{\GeV}$, $m_W = \unit{80.385}{\GeV}$, $m_H = \unit{125.1}{\GeV}$, $\alpha_s(m_Z) = 0.118$ \cite{Tanabashi:2018oca}.

\begin{figure}[t!]
\includegraphics[width=0.4\textwidth]{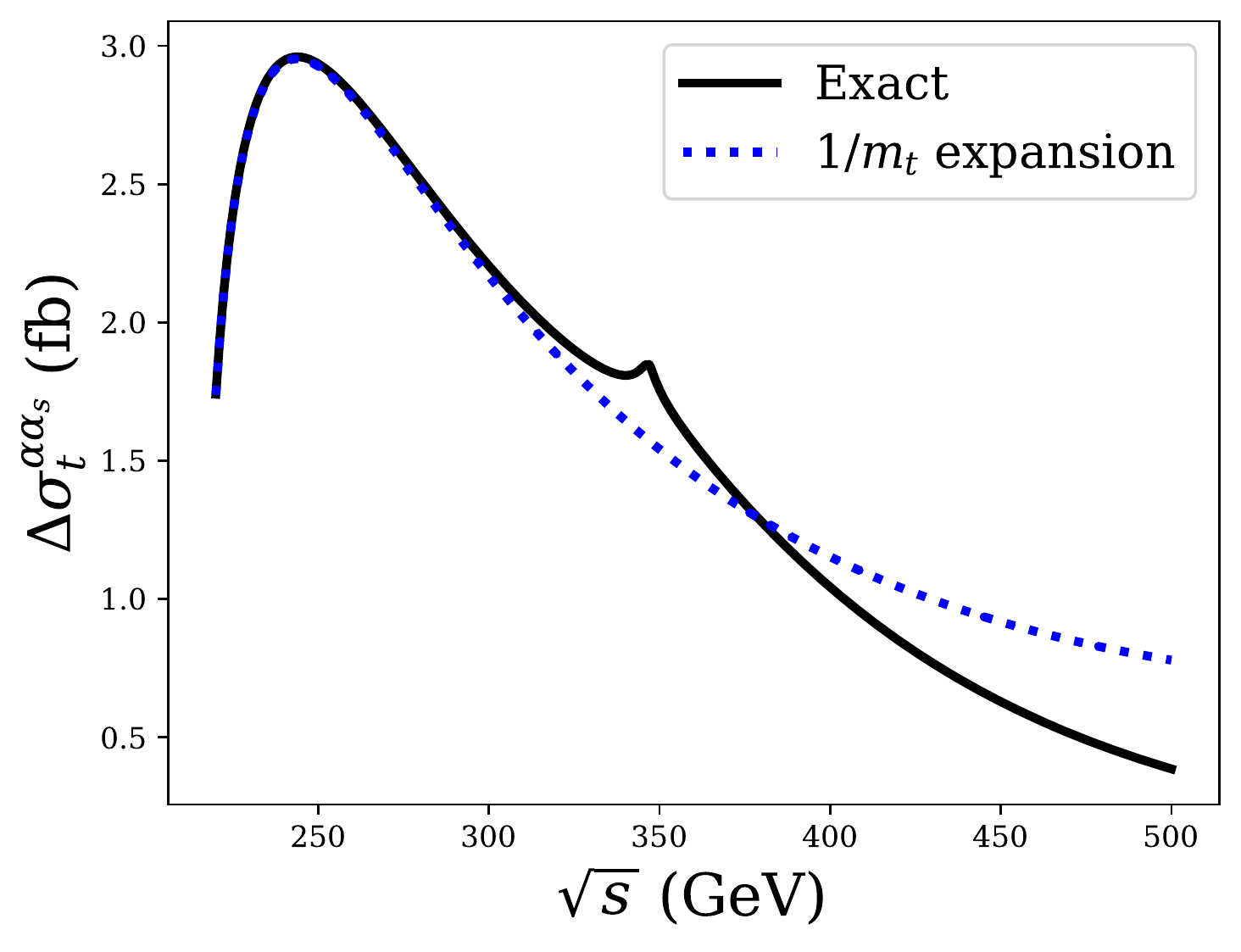}
\caption{\label{fig:exact_vs_expansion}NNLO $\mathcal{O}(\alpha\alpha_s)$ corrections from top quark loops to the $e^+e^- \to ZH$ production cross sections.}
\end{figure}

In Fig.~\ref{fig:exact_vs_expansion}, we show the exact top quark loop contributions to the $e^+e^- \to ZH$ cross sections for a center-of-mass energy $\sqrt{s}$ ranging from $\unit{220}{\GeV}$ to $\unit{500}{\GeV}$. Thanks to our analytic results, it is much faster to perform the numeric computation compared to purely numeric methods such as sector decomposition. Especially for the $t\bar{t}$ threshold region $\sqrt{s} \sim 2m_t$, the purely numeric integration is badly convergent, while it poses no difficulty for our analytic formulas. In the plot we also show the result of large $m_t$ expansion derived in \cite{Gong:2016jys} up to $\mathcal{O}(1/m_t^4)$. As expected, the expansion behaves well for low energies, but ceases to be valid near or above the $t\bar{t}$ threshold.

\begin{figure}[t!]
\includegraphics[width=0.4\textwidth]{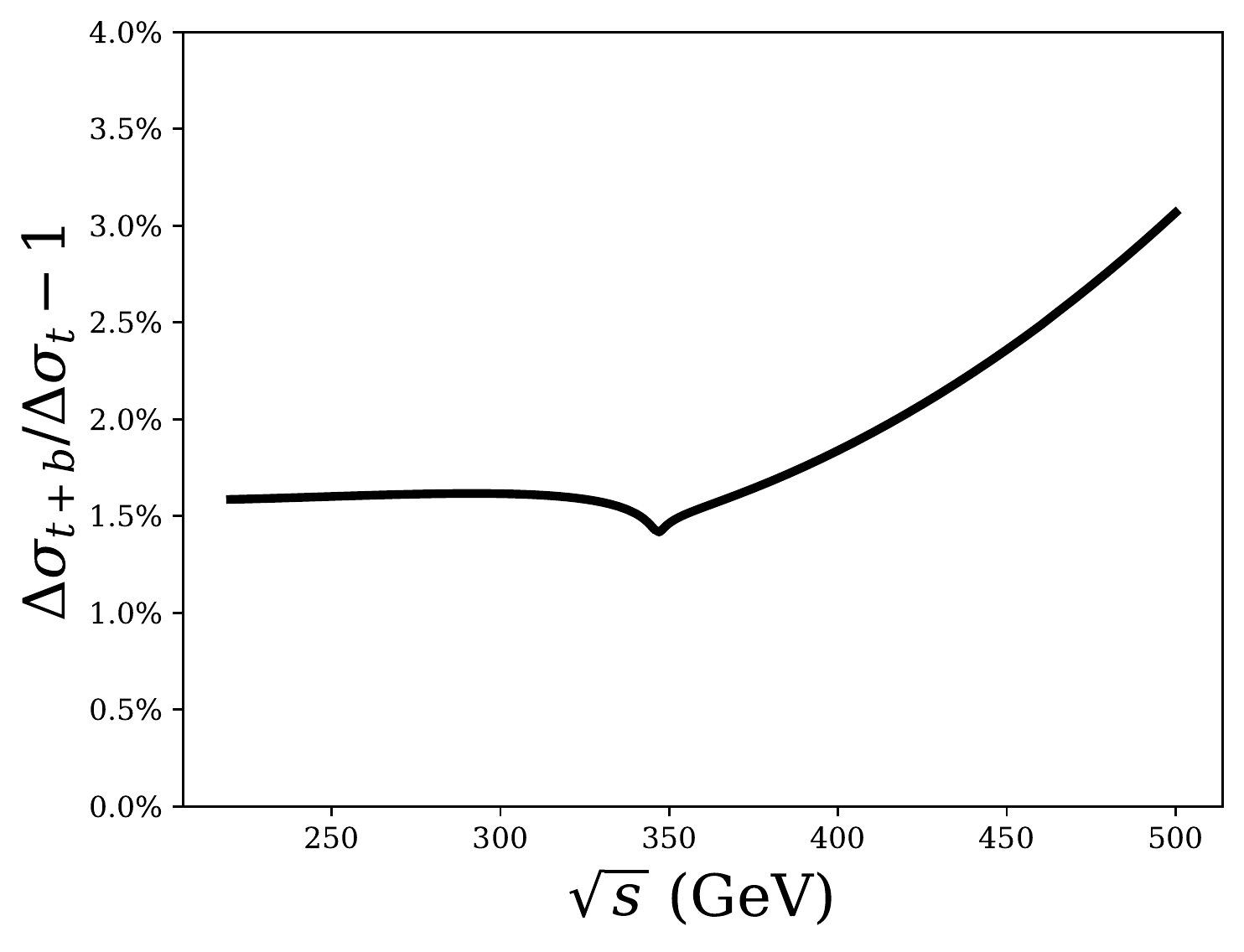}
\caption{\label{fig:b_vs_nob}Relative corrections from bottom quark loops to the $e^+e^- \to ZH$ production cross section.}
\end{figure}

In Fig.~\ref{fig:b_vs_nob}, we show the impact of adding the contributions from bottom quark loops to the $ZH$ cross section. Again, computing this contribution is rather time-consuming with sector decomposition, but is much faster with the analytic results at hand. Phenomenologically this contribution only amounts to a few percent of the $\mathcal{O}(\alpha\alpha_s)$ corrections (which is below 1 per mille of the total cross section), and is therefore not important.

\begin{figure}[t!]
\includegraphics[width=0.4\textwidth]{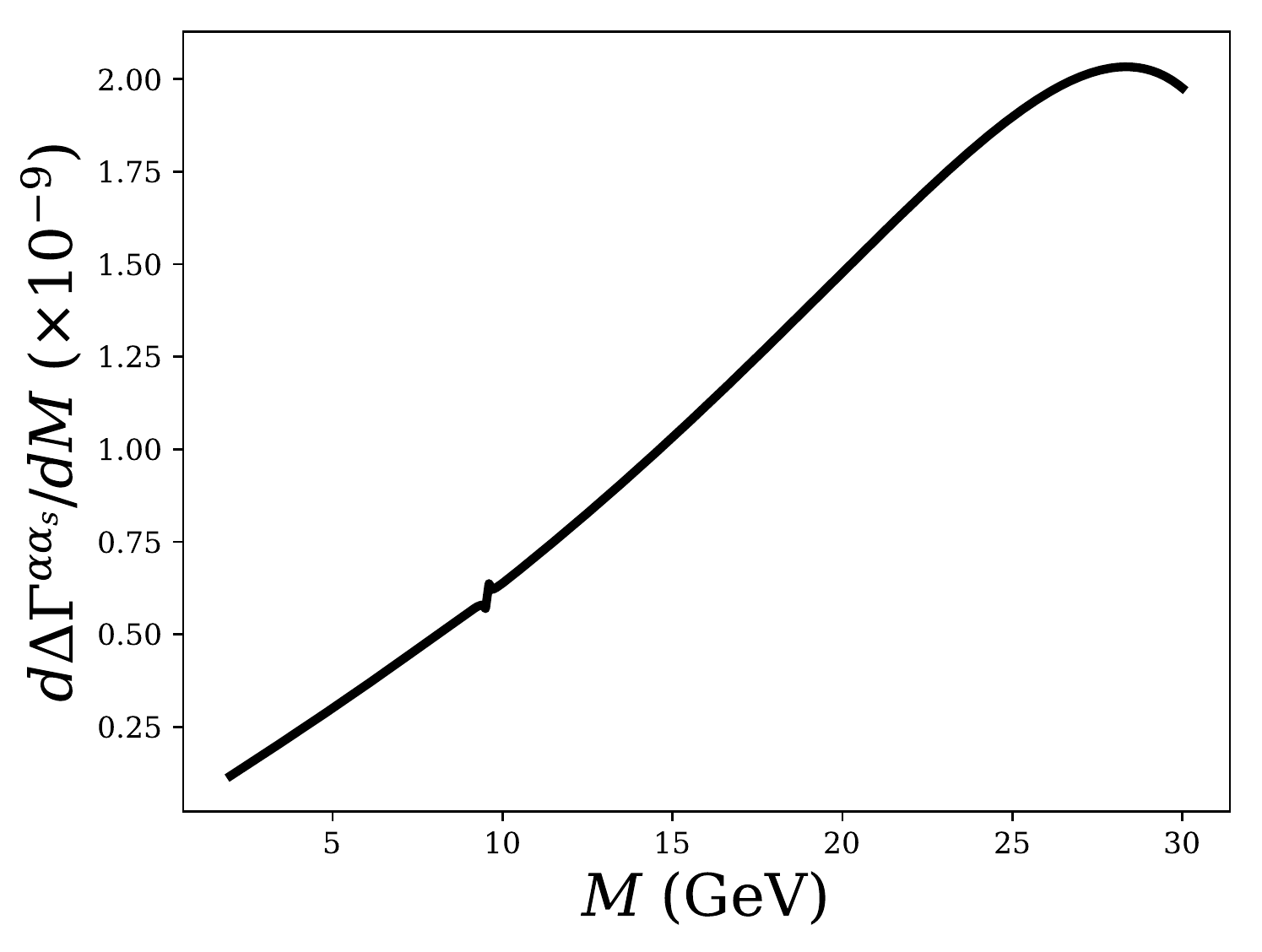}
\caption{\label{fig:decay}NNLO $\mathcal{O}(\alpha\alpha_s)$ corrections to the $H \to Zl^+l^-$ differential decay rate as a function of the lepton pair invariant mass $M$, including both top and bottom quark contributions.}
\end{figure}

We now turn to the leptonic decay $H \to 4l$. The leading $m_t^2$ enhanced contributions at $\mathcal{O}(\alpha\alpha_s)$ have been considered in \cite{Kniehl:2012rz}. Here we give the result for the exact $\mathcal{O}(\alpha\alpha_s)$ corrections including bottom quark loops. For simplicity, we consider the process $H \to Zl^+l^-$ and treat the leptons as massless. A more dedicated study, including the decay of both $Z$ bosons and the lepton mass effects, will be presented in \cite{future}. In Fig.~\ref{fig:decay} we show the $\mathcal{O}(\alpha\alpha_s)$ corrections to the differential decay rate $d\Gamma/dM$, where $M$ is the invariant mass of the lepton pair. We incorporate both top quark and bottom quark loop contributions. Note in particular the kink at $M \approx 2m_b$, which is due to the Coulomb singularity at the $b\bar{b}$ threshold. A proper treatment of this region would require resumming the Coulomb exchanges as well as dealing with non-perturbative effects, which is beyond the scope of this work.

\section{Summary and outlook}

In this paper, we calculate analytically the two-loop triangle integrals entering the $\mathcal{O}(\alpha\alpha_s)$ corrections to the $HZV$ vertex. We derive the canonical-form differential equations for the 41 master integrals appearing in the calculation. For integrals with 4 mass scales, these differential equations are not easy to solve due to the presence of many non-rational functions. We are able to find fully analytic solutions up to weight 3 in terms of multiple polylogarithms. We apply our results to the $e^+e^- \to ZH$ production cross section and the $H \to ZZ^* \to 4l$ decay width, including both top quark loops and bottom quark loops. For the bottom quark loop contributions, and for cases when the collider energy is near the $t\bar{t}$ threshold, the integrals are rather time-consuming using purely numeric methods such as sector decomposition. This poses no difficulty for our analytic results, whose numeric evaluation is efficient for all phase-space points.

Loop integrals with many mass scales are very common in electroweak physics, Higgs physics and top quark physics. However, it is not easy to evaluate them in closed form, especially at high orders in $\epsilon$. Our result serves as a prototype to study the analytic structure of multi-loop multi-scale Feynman integrals. For the $HZV$ vertex, it is interesting to study its behaviors in various asymptotic limits. For example, near the $t\bar{t}$ threshold $\sqrt{s} \sim 2m_t$, it is expected that the amplitude can be factorized in the framework of non-relativistic effective field theory \cite{Beneke:2003xh, Beneke:2004km, Grober:2017uho}. It is interesting to see how such a factorization is achieved at $\mathcal{O}(\alpha\alpha_s)$ using our analytic results. Another interesting region is the small internal mass limit, which is relevant for bottom quark loops. In this limit, logarithms of the internal mass may develop at each order in perturbation theory, which have been studied in \cite{Liu:2017vkm, Liu:2018czl} for the $Hgg$ vertex. It is interesting to investigate the same limit for the $HZV$ vertex, where more scales come into play. These studies will be presented in a forthcoming article \cite{future}.

Finally, although we have obtained the analytic results at weight-3 in terms of MPLs, their expressions are rather lengthy, mainly due to the variable change Eq.~\eqref{eq:tv}. It is possible that we can find a suitable basis of trilogarithm functions to shorten the expressions. This may also help us to construct an analytic form for the weight-4 part of the solution. It is interesting to investigate these in the future.

\vspace{1ex}

{\em Acknowledgments:\/}
This work was supported in part by the National Natural Science Foundation of China under Grant No. 11575004 and No. 11635001.

\end{document}